\documentclass[12pt]{article}
\input{wasyfont}

\def\Journal#1#2#3#4{{#4} {\it #1} {\bf #2}, #3 }
\def\ac{\overline{\alpha}}
\def\bc{\overline{\beta}}
\def\gc{\overline{\gamma}}
\def\dc{\overline{\delta}}
\def\ec{\overline{\epsilon}}
\def\rc{\overline{\rho}}
\def\mc{\overline{\mu}}
\def\tc{\overline{\tau}}
\def\pc{\overline{\pi}}
\def\kc{\overline{\kappa}}
\def\nc{\overline{\nu}}
\def\s5{\sqrt{5}}
\def\D{\Delta}

\def\b{\beta}
\def\g{\gamma}

\def\d{\delta}
\def\e{\epsilon}
\def\t{\theta}

\def\k{\kappa}
\def\l{\lambda}
\def\r{\rho}
\def\s{\sigma}

\def\bar{\overline}

\newcommand{\vrk}{\hfill $\Box$}
\newcommand{\vs}{\vspace{.1cm}}
\newcommand{\vvs}{\vspace{.3cm}}

\begin{document}

\begin{center}

{\bf Note on aligned Petrov type D purely magnetic perfect fluids}
\vvs

Lode Wylleman\footnote{Research assistant supported by the Fund for
Scientific Research Flanders(F.W.O.), e-mail:
lwyllema@cage.ugent.be} and Norbert Van den Bergh\footnote{e-mail:
norbert.vandenbergh@ugent.be}

\vvs

%\address
{Faculty of Applied Sciences TW16, Gent University, Galglaan 2, 9000 Gent, Belgium}

\end{center}
\vvs

\begin{abstract} We recently showed (Class. Quantum Grav.\ 23, 3353;
gr-qc/0602091) that aligned Petrov type D purely magnetic perfect
fluids are necessarily locally rotationally symmetric (LRS) and
hence are all explicitly known. We provide here a more transparent
proof.\end{abstract}

%\pacs{0420}


%\maketitle

Aligned purely magnetic perfect fluids are defined to be
non-conformally flat solutions of Einstein's field equations
\begin{equation}
G_{ab}\equiv R_{ab}-\frac{1}{2}R g_{ab}= (w+p) u_a u_b -p g_{ab},
\end{equation}
for which the electric part of the Weyl tensor w.r.t.\ to the unique
normalized timelike eigenvector field $u^a$ of the Ricci tensor
vanishes:
\begin{equation}\label{E_ab}
E_{ac}\equiv C_{abcd} u^b u^d=0.
\end{equation}
All locally rotationally symmetric solutions of this kind were
recently found \cite{Lozanovski2,Lozanovski3}. They are of class III
or I in the classification by Stewart and Ellis \cite{StewartEllis},
wherefor the metrics can then be explicitly constructed or
determined up to a single third-order differential equation,
respectively. It was shown in \cite{VdBWyll}, within the
Newman-Penrose formalism, that these LRS classes actually exhaust
Petrov type D. To come to the result, use was made of a further
tetrad fixation, which is unnecessary and hides somewhat the exact
freedom which is left in the NP and Bianchi-equations at a certain
level of the reasoning. The streamlined proof below is designed to
bring more clarity, and also demonstrates in a straightforward way
that the LRS classes are I or III.
%, without relying on the general LRS metric forms (as was done in \cite{Lozanovski3}).
%This also
%stresses on the duality, in some sense,  between solutions in both
%classes.
We will still make use of the Newman-Penrose formalism and follow
the notation of \cite{Kramer}, whereby the Newman-Penrose equations
(7.21a -- 7.21r) and Bianchi identities (7.32i -- 7.32h) will be
indicated as ($np1$ -- $np18$) and ($b9$ -- $b11$). Regarding (7.32a
-- 7.32h), appropriate combinations with (7.32i -- 7.32h) are made
to remove derivatives of the Ricci scalar, and the resulting
equations are ($b1$ -- $b11$). In the philosophy of \emph{not}
further fixing ${\cal B}$, the GHP `edth' operator $\dh$ will also
come into play.

We use a canonical type D tetrad ${\cal B}$, with
\begin{equation}\label{e:psis} \Psi_0=\Psi_1=\Psi_3=\Psi_4=0 ,\end{equation} with
 the condition (\ref{E_ab}) being expressed as
\begin{equation}\label{e:psi2}
\overline{\Psi_2}=-\Psi_2\neq 0
\end{equation}
Choosing a boost in the $(k,\ell)$ plane such that the fluid
velocity $u=(k+\ell)/ \sqrt{2}$ and introducing $S=w+p$ as a new
variable, one has
\begin{equation}\label{e:phis}
\Phi_{00}=\Phi_{22}=2 \Phi_{11}=\frac{S}{4}\ \textrm{and}\ R=4 w-3
S.
\end{equation}
\vs

{\bf Theorem.} A PMpf of Petrov type D is LRS class I or III. More
precisely, the spin coefficients of a canonical Weyl null-tetrad
obey $\l=\s=\b+\ac=\k=\tau=\pi=\nu=0$, with moreover
\begin{equation}\label{extra}
    \mu=q\,\rho,\quad g=q\,e, \Psi_2=\frac{q}{2}(\rc-\r)(2e+\r+\rc),
\end{equation}
where $g\equiv \g+\gc$ and $e\equiv \e+\ec$, and $q$ is 1 for class
I and -1 for class III.\\
%\begin{itemize}
%\item $\mu=\rho$, $g=e$ for class I,
%\item $\mu=-\rho$, $g=-e$ for class III,
%\end{itemize}
%where $g\equiv \g+\gc$ and $e\equiv \e+\ec$.
% and the following algebraic
%relation in $\Psi_2,S$ and the remaining spin coefficients holds:
%\end{theorem}

\textsc{Proof.} %The relation (\ref{e:Rextra}) immediately follows
%from $\Re((bi1)+(bi2))$
From $(bi5)$ and $(bi6)$ one gets
\begin{eqnarray}
    (D-\D)S &=& -(g+e)S-9(\r+\mu-(\rc+\mc))\Psi_2,\label{d:DminDeltaS}\\
    (D-\D)\Psi_2 &=&
    \frac{3}{2}(\r+\mu+\rc+\mc)\Psi_2-\frac{1}{4}(\r+\mu-(\rc+\mc))S,
    \label{d:DminDeltaPsi2}\\
    (D+\D)\Psi_2 &=& -\frac{3}{2}(\mu-\rc+(\mc-\r))\Psi_2,
    \label{d:DplusDeltaPsi}
\end{eqnarray}
together with an algebraic relation
\begin{equation}\label{e:Rextra}
    18(\rho-\rc+\mc-\mu)\Psi_2-
(\rho+\rc-\mu-\mc+2(e-g))S=0.
\end{equation}
 Next, $\l=\s=0$ immediately follows from $(bi2)$ and $(bi3)$. From
$(bi1)+\bar{(bi4)}+\bar{(bi7)}+(bi8)$ one gets
\begin{equation} \kappa+3(\overline{\pi}+\tau)
+\overline{\nu} = 0. \label{e:lsnu}
\end{equation}
The first key observation is that herewith, the combination
$(np2)+3\overline{(np4)} +3 (np10) + \overline{(np14)}$ is algebraic
and factorises as follows:
\begin{equation}\label{e:main}
(\b+\ac)(2\k+3 \tau+3 \pc)=0.
\end{equation}
If $\b+\ac$ were not identically zero, then from (\ref{e:lsnu}) and
(\ref{e:main}) one would obtain
%\begin{equation}
$\kappa = \nc = -\frac{3}{2} (\tau+\pc)$
%\end{equation}
after which $(bi1)-\bar{(bi4)}$ and $(np2)-\bar{(np10)}$
respectively yield
\begin{equation}
\b+\ac = \tau+\pc,\quad (\b+\ac)(\tau+\pc)=0,
\end{equation}
such that one would come to the contradiction $\b+\ac=0$.  With
$\b+\ac=0$, we respectively get from
$(np11),\bar{(np13)},\bar{(np4)}+(np5)$ and $(np15)+\bar{(np18)}$
that
\begin{eqnarray}
 \d\r&=&\t(\r-\rc), \label{d:rho} \\
 \d\mc&=&\pc(\mu-\mc)-(\k+3(\t+\pc))(\r-\rc),  \\
  \d e&=&-\pc(\rc+\e)+\k(\mu+g),  \\
  \d g&=&\t(\mu+g)+(\k+3(\t+\pc))(\rc+e) \label{d:g},
\end{eqnarray}
whereas from $(bi1)$, $(bi8)-\bar{(bi7)}$ and
$(np2),\bar{(np7)},(np16)$ we get an autonomous dynamical system in
$\Psi[2],S,\k,\tau,\pc$ with $\dh$ as differential operator, for
which (a) $\dh\Psi_2$ and $\dh S$ are linear and homogeneous in
$\Psi_2$ and $S$, and (b) $\k,\tau,\pc$ form a subsystem:
\begin{eqnarray}
  \dh\Psi_2 &=& \frac{3}{2}(\tau-\pc)\Psi_2 -\frac{1}{2}(\tau+\pc)S=\d\Psi_2,\label{d:edthPsi2}\\
  \dh S  &=& -12\k\Psi_2+(\k-\pc)S=\d S,\label{d:edthS}\\
  \dh\k &=& (\tau-\pc)\k, \label{d:edthkc}\\
  \dh\pc &=& -(3\k+\pc)\pc-(\k+3\tau)\k, \label{d:edthpi}\\
  \dh\tau &=& (3\k+\tau)\tau+(\k+3\pc)\k.\label{d:edthtc}
\end{eqnarray}
However, from $\bar{(bi4)}-(bi1)$ we get that these variables are
constrained by
\begin{equation}\label{e:edthconstraint1}
    3(2\k+3\tau+3\pc)\Psi_2+(\tau+\pc)S=0.
\end{equation}
Hence, on applying $\dh$ to (\ref{e:edthconstraint1}), the
homogeneous and linear character in $\Psi_2$ and $S$ is preserved,
and as $(\Psi_2,S)\neq (0,0)$ we must have
\begin{equation}\label{e:edthconstraint2}
    2\k^2+9\tau^2+6\pc^2+7\k\tau+5\k\pc+15\tau\pc=0.
\end{equation}
Applying $\dh$ to \ref{e:edthconstraint2} one further finds
\begin{equation}\label{e:edthconstraint3}
    2\k^3+18\tau^3-12\pc^3+13\tau\k^2+23\k\tau^2+5\pc\k^2-\k\pc^2
    +15\pc\tau^2-15\tau\pc^2+16\tau\k\pc=0.
\end{equation}
Eliminating $\pc$, resp.\ $\t$, from (\ref{e:edthconstraint2}) and
(\ref{e:edthconstraint3}) by calculating resultants one obtains
\begin{equation}\label{e:edthresultant}
    %(27\pi^2+44\pi\tc+27\tc^2)*(\tc+\pi)^4=0, \quad
    (9\k^2-2\k\tau+9\tau^2)\k^2(\k+\tau)^2=0, \quad
    (3\k^2+4\k\pc+3\pc^2)\k^2(\k-\pc)^2=0.
\end{equation}

Suppose $\k\neq 0$. By substituting into (\ref{e:edthconstraint2})
and (\ref{e:edthconstraint3}) the relations $\k=-\tau$ and $\k=\pc$
one finds that these are equivalent, hence $\tau+\pc=0$, in
contradiction with (\ref{e:edthconstraint1}) and $\Psi_2\neq 0\neq
\k$. Hence possible
ratio's $\t/\k$ and $\pc/\k$ %$\frac{\tau}{\k}$ and $\frac{\pc}{\k}$
are roots of the first factors in (\ref{e:edthconstraint3}) divided
by $\k^2$. Checking (\ref{e:edthconstraint2}) for the four different
possibilities, one finds that only two couples are allowed:
\begin{equation}\label{e:soltaupc}
    %(\tau,\k) = \left(-\frac{\pi}{27}(22+\e\, 7
    %i\sqrt{5}),\frac{\pi}{3}(-2+\e\,
    %i\sqrt{5}\right),\quad \e=\pm 1.
    %\left(\frac{\tau}{\k},\frac{\pc}{\k}\right)=\left(\frac{1}{9}(1+z 4i\sqrt{5}),
    %-\frac{1}{3}(2+z i\sqrt{5})\right),\quad z=\pm 1.
    \left(\tau,\pc\right)=\left(\frac{1}{9}(1+z 4i\sqrt{5})\k,
    -\frac{1}{3}(2+z i\sqrt{5})\k\right),\quad z=\pm 1.
\end{equation}
Inserting this in (\ref{e:edthconstraint1}) and dividing by $\k$ we
correspondingly find $S$ and $\Psi_2$ to be constantly proportional:
\begin{equation}\label{e:solS}
    S=z \frac{9i}{\sqrt{5}}\Psi_2.
\end{equation}
We immediately get two additional (real) algebraic equations in
$e,g,\r,\rc,\mu,\mc$ from (\ref{e:solS}), by substituting it into
(\ref{e:Rextra}) on one hand, and applying $D-\D$ to it and use
(\ref{d:DminDeltaS})-(\ref{d:DminDeltaPsi2}). With $\Psi_2\neq 0$,
this respectively yields
\begin{eqnarray}
  z\sqrt{5}(\mu+\mc-(\r+\rc)+2(g-e))+10i(\mu-\mc-(\r-\rc)) &=& 0, \label{e:eq1} \\
  z2\sqrt{5}(3(\mu+\mc+\r+\rc)+2(g+e))-11i(\mu-\mc+\r-\rc) &=& 0. \label{e:eq2}
\end{eqnarray}
To come to a quick contradiction we would like to have four more
real (or two more complex) homogeneous and linear equations in
$e,g,\r,\rc,\mu,\mc$. The amazing fact is that \emph{just} two such
complex equations (and not more!) become easily available, as
follows. Combining (\ref{e:soltaupc}) and (\ref{e:lsnu}), one finds
that the ratio's $\nc/tau$ and $\pc/\k$ are equal
\begin{equation}\label{e:equalratios}
\frac{\nc}{\tau}=\frac{\pc}{\k}=-\frac{1}{3}(2+z i\sqrt{5}):=a(z).
\end{equation}
The second key observation is that herewith, $\bar{(np9)}-a(z)(np3)$
becomes algebraic and of the wanted form:
\begin{equation}\label{e:eq3}
    z\sqrt{5}(\r-\mc+6(g-e))-i(5(\r+\mc)+12(g+e))=0.
\end{equation}
By means of (\ref{d:rho})-(\ref{d:g}) we can get another algebraic
equation on applying $\d$ on (\ref{e:eq3}), but this turns out to
give an identity when inserting (\ref{e:soltaupc}). However, we can
build one other homogeneous and linear complex equation in
$\r,\mc,e,g$ by eliminating $\rc$ and $\mu$ from
(\ref{e:eq1}),(\ref{e:eq2}) and the complex conjugate of
(\ref{e:equalratiosnp}). This yields
\begin{equation}\label{e:kraak}
    z\sqrt{5}(25(\r-\mc)+31(g-e))+2i(65(\r+\mc)+63(g+e))=0.
\end{equation}
Applying $\d$ on this, inserting (\ref{e:soltaupc}) and dividing by
$\k$ now gives
\begin{eqnarray}\label{e:eq5}
  &&  z(255\rho+873\rc-13\mu+765\mc+752g+1128e)\nonumber\\
  &&-i\sqrt{5}(357(\r-\rc)+529\mu-153\mc+376g)=0.
\end{eqnarray}
Thus
$\{(\ref{e:eq1}),(\ref{e:eq2}),(\ref{e:eq3}),\bar{(\ref{e:eq3})},
(\ref{e:eq5}),\bar{(\ref{e:eq5})}\}$ forms a homogeneous and linear
system in $\r,\rc,\mu,\mc,e,g$, the determinant of which is computed
to be non-zero (and independent of $z$). Thus the only solution is
the zero-solution. However, inserting $\r=\mu=0$ in the imaginary
part of $(np12)$ then yields $\Psi_2=0$: contradiction.

Hence $\k=0$. Substituting this into (\ref{e:edthconstraint2}) and
(\ref{e:edthconstraint3}) one finds
\begin{equation}\label{plop}
    (3\t+2\pc)(\tau+\pc)=(-6\tau^2+\pc\t+4\pc^2)(\t+\pc)=0
\end{equation}
such that $\t+\pc=0$, and $\nu=0$ by (\ref{e:lsnu}). With the so far
obtained specifications, $(bi10)$ reduces to $\d w=\dc w=0$, while
from $(bi9)$ and $(bi11)$ we respectively get:
\begin{eqnarray}
  Dw &=& \frac{9}{2}(\r+\mu-(\rc+\mc))\Psi_2+\frac{1}{2}(g-e+\r+\rc-\mu-\mc)S, \\
  \D w &=&
  -\frac{9}{2}(\r+\mu-(\rc+\mc))\Psi_2+\frac{1}{2}(g-e+\r+\rc-\mu-\mc)S.
\end{eqnarray}
Herewith, the $[\d,\dc]$ commutator applied on $w$ is algebraic, and
elimination of $\Psi_2$ from it by combination with (\ref{e:Rextra})
yields a factorisation
\begin{equation}\label{e:LRSIorIII}
    S(\mu-\rc-(\mc-\r))(\mu-\rc+(\mc-\r))=0.
\end{equation}
Hence $\mu-\rc$ is either real or imaginary, where the real case
corresponds to vanishing vorticity of the $u^a$-congruence, see the
appendix. Application of the $[\d,D+\D]$ commutator to $w$ results
for both cases in
\begin{equation}\label{e:taufactor}
    \tau(\mu-\rc+g-e)=0.
\end{equation}
In the case where $\mu-\rc$ is real, the imaginary part of $(np12)$
precisely gives
\begin{equation}\label{e:Psi2}
    \Psi_2=\frac{1}{2}(\r-\rc)(\mu-\rc+g-e)
\end{equation}
such that the second factor of \ref{e:taufactor} cannot vanish. It
can vanish neither when $\mu-\rc$ is imaginary, as then $\mu-\rc$
would be zero as $g-e$ is real. Therefore $\tau,\k,\nu,\pi,\l$ and
$\s$ are all zero, and hence the solutions are LRS according to
corollary 1 in \cite{GoodeWainwright}.

Further, the Ricci equations $(np1),(np8),(np17)$ and $(np14)$
respectively yield
\begin{eqnarray}
  D\r &=& \r(\r+e)+\frac{S}{4} \\
  D\mu &=& \mu(\rc-e)+\Psi_2+\frac{w}{3}-\frac{S}{4}\\
  \D\r &=& -\r(\mc-g) -\Psi_2-\frac{w}{3}+\frac{S}{4}\\
  \D\mu &=& -\mu(\mu+g)-\frac{S}{4}.
\end{eqnarray}

The non-rotating case ($\mu-\rc$ real) was fully treated in
\cite{Lozanovski2}. Application of the commutator $[\d,\dc]$ to
$\Psi_2$ and the evolution operator $D+\D$ to $\mu+\r-\mc-\rc=0$
respectively yield
\begin{equation}\label{e:plok}
    \Psi_2(\r-\rc)(\r+\mu)=0,\quad (\r-\rc)(e+g)=0,
\end{equation}
such that $\r+\mu=g+e=0$ by (\ref{e:Psi2}), where $g+e=0$ is
equivalent to the perfect fluid $u^a$-congruence being geodesic, see
the appendix. As $\rho=\mu=0$ is not allowed by (\ref{e:Psi2}), one
has $\rho/\mu=-1 <0$, and the corresponding space-times are LRS
class III, see \cite{GoodeWainwright}.

For imaginary $\mu-\rc\neq 0$, application of $D-\D$ to
$\r+\rc-\mu-\mc=0$ yields
\begin{equation}\label{e:jiha1}
    (\mu+\mc)(e-g)+(\mu-\r)(\mu-\rc)=0,
\end{equation}
whereas from the application of the commutator $[\d,\dc]$ to
$\Psi_2$ one now obtains
\begin{equation}\label{e:aha1}
    (\mu-\rho)[(\r-\mc)S+6(\mu+\mc)\Psi_2]=0.
\end{equation}
If the second factor in (\ref{e:aha1}) vanishes, then eliminating
$S$ from it in combination with (\ref{e:Rextra}) yields
\begin{equation}\label{e:jiha2}
    (\mu+\mc)(e-g)+3(\mu-\r)(\mu-\rc)=0.
\end{equation}
Hence, consistency with (\ref{e:jiha1}) requires $\r=\mu$ anyway.
Substitution of this into (\ref{e:aha1}) finally gives $g=e$, and
the $u^a$-congruence is shear-free and non-expanding, see the
appendix. The imaginary part of $(np12)$ gives the expression
(\ref{extra}) for $\Psi_2$ with $q=1$. Again, $\rho=\mu=0$ is not
allowed, hence $\rho/\mu=1>0$ and the corresponding space-times are
LRS class I \cite{GoodeWainwright}. \vrk\\

{\bf Remark.} All LRS class I, resp.\ class III, solutions are
stationary, resp.\ orthogonally spatially homogeneous, and hence
exhibit an equation of state, which can be most explicitly be
determined for LRS class III PMpf's, see \cite{Lozanovski2}. LRS
class II only contains purely electric solutions. A beautiful
overview of LRS space-times in a 1+3 covariant approach, summarizing
many features of the different classes and discussing their
practical meaning, can be found in \cite{vanElst1}.

\appendix

\section*{Appendix}

Choosing an orthonormal tetrad such that $\delta\equiv(e_1 - i
e_2)/\sqrt{2}$, $D \equiv(e_3+e_4)/\sqrt{2}$ and $\Delta \equiv
(e_4-e_3)/\sqrt{2}$ ($e_4=u$ being the fluid velocity), the
components of the fluid kinematical quantities are given by the
following expressions:

(expansion tensor)
\begin{eqnarray}
\theta_{12} = 0 \\
\theta_{13}+i \theta_{23} = (\alpha+\bc+2\pi+2\tc)/\sqrt{2} \\
\theta_{11} = \theta_{22} = (\mu+\mc-\rho-\rc)/(2\sqrt{2})\\
\theta_{33}= (\epsilon+\ec-\gamma-\gc)/\sqrt{2}
\end{eqnarray}

(acceleration vector)
\begin{eqnarray}
\dot{u}_1+ i \dot{u}_2 = -\sqrt{2}(\pi+\kc+2\tc)\\
\dot{u}_3 = (\epsilon+\ec+\gamma+\gc)/\sqrt{2}
\end{eqnarray}

(vorticity vector)
\begin{eqnarray}
\omega_1+i \omega_2 = \frac{i}{2} (\alpha+\bc - 2 \tc -2 \pi)/\sqrt{2}\\
\omega_3 = \frac{i}{2}(\rho-\rc+\mu-\mc)/\sqrt{2}
\end{eqnarray}

%\section*{References}

\end{document}